\documentclass[11pt]{article}

\usepackage[T1]{fontenc}
\usepackage{lmodern}
\usepackage{amsmath,amssymb,bm}
\usepackage{authblk}
\usepackage{cite}
\usepackage[margin=0.88in]{geometry}
\usepackage{graphicx}
\usepackage{flafter}
\usepackage{microtype}
\usepackage[colorlinks=true,citecolor=blue,linkcolor=blue,urlcolor=blue]{hyperref}
\usepackage{color}
\usepackage{ulem}

\newcommand{\cfin}{c_{\mathrm{fin}}}
\newcommand{\Ba}{B_{\alpha}}
\newcommand{\ella}{\ell_{\alpha}}
\newcommand{\Lb}{L_b}
\newcommand{\Ls}{L_s}

\begin{document}

\title{Charged soliton--black hole phase transitions in
three-dimensional Einstein--Gauss--Bonnet gravity}

\author[1]{Cendikiawan Suryaatmadja\thanks{csuryaat@uwaterloo.ca}}
\author[1,2]{Robert B. Mann\thanks{rbmann@uwaterloo.ca}}

\affil[1]{Department of Physics and Astronomy, University of Waterloo,
Waterloo, Ontario N2L 3G1, Canada}
\affil[2]{Perimeter Institute for Theoretical Physics, Waterloo,
Ontario N2L 2Y5, Canada}

\date{\today}
\maketitle

\begin{abstract}
We construct a magnetic soliton by double Wick rotating the static
electrically charged black hole in three-dimensional
Einstein--Gauss--Bonnet gravity.  We match the boundary metrics and Maxwell
sources under the chosen boundary condition and compute the difference of the renormalized Euclidean actions.  Because the Maxwell potential grows
logarithmically at infinity, the chosen finite Maxwell boundary term changes
both the fixed Maxwell source and the renormalized action.  The charge contributions for both the soliton and the black hole appear with plus signs in $\Delta I^{(\cfin)}$.  For the grand-canonical boundary condition $\cfin=1$, the black-hole and soliton branches can cross with different slopes.  For the example considered below, the black-hole branch satisfying the response conditions ends at a maximum temperature, and the coexistence curve has a $T\to0^+$ endpoint at two opposite values of the spatial Maxwell source.  At a nonextremal crossing, the entropy difference is $S_b-S_s=S_b$.  For these two static solutions, equality of both current components requires both charges to vanish.
\end{abstract}

\section{Introduction}

Compactifying one direction of a planar anti--de Sitter black hole permits a
second smooth filling of the same boundary torus: the AdS soliton
\cite{HorowitzMyers1998}.  The two fillings differ in which boundary circle is
contractible, and  a comparison of their Euclidean actions 
indicates that phase transitions between the soliton and the black hole can take place \cite{SuryaSchleichWitt2001}.  More recently, the corresponding problem 
for the charged case in  Einstein--Maxwell gravity in four dimensions was studied~\cite{AnabalonEtAl2022}, where the coexistence surface depends on the
electric potential and magnetic flux.

In four-dimensional Einstein gravity coupled to nonlinear electrodynamics,
Quijada et al.~\cite{QuijadaEtAl2024} found reentrant phase transitions
and triple points at which two magnetic solitons coexist with an
electrically charged planar AdS black hole. Bravo-Gaete et
al.~\cite{BravoGaeteEtAl2026} subsequently constructed multicritical
points involving several magnetic solitons and a single electrically
charged black hole, using a polynomial structural function in
Pleba{\'n}ski nonlinear electrodynamics.

{In (2+1)-dimensional Einstein gravity the neutral soliton is global
AdS$_3$~\cite{SuryaSchleichWitt2001}.  Its Euclidean section and that of the
nonrotating BTZ black hole~\cite{BanadosTeitelboimZanelli1992} are related by
a coordinate transformation that exchanges the two boundary circles, so the
transition between them occurs when the two circles have equal length.  This
is the Hawking--Page transition~\cite{HawkingPage1983} between the BTZ black
hole and thermal AdS$_3$~\cite{MaldacenaStrominger1998,EuneKimYi2013}.  The
same exchange relates the neutral black hole and soliton of the
Einstein--Gauss--Bonnet theory considered below (Sec.~\ref{sec:alpha0}).  With a Maxwell field, the magnetically
charged soliton is obtained only after analytically continuing the charge.  No real
coordinate transformation relates it to the electrically charged black hole,
since $F_{ab}F^{ab}$ has opposite signs for the two.
Here we demonstrate that phase transitions can take place between
electrically charged black holes and their magnetically charged soliton
counterparts in (2+1)-dimensional Einstein--Gauss--Bonnet gravity, or 3DEGB.
In the reduced variables of Sec.~\ref{sec:phase}, the resulting phase
diagram does not depend on the Gauss--Bonnet coupling, which only sets the
effective AdS radius used to convert to physical units.  The same transitions
therefore occur in (2+1)-dimensional Einstein--Maxwell gravity
(Sec.~\ref{sec:alpha0}).}

 The static
Einstein--Maxwell magnetic solution was obtained from the electrically
charged BTZ solution by analytic continuation~\cite{CataldoSalgado1996}.  Auzzi et al. studied this background as an
AdS--Melvin soliton obtained by double analytic continuation and analyzed its
logarithmic Maxwell data.  Their model also contains a charged scalar, and the
boundary condition used there fixes the logarithmic coefficient
\cite{AuzziEtAl2025}.

Here we study the corresponding EGB soliton with a
nonzero Maxwell field and compare it with the electrically charged black hole
at fixed boundary metric and renormalized Maxwell source.
The static electrically charged black hole we consider was obtained in the original formulation of 3DEGB~\cite{HennigarKubiznakMann2021}.  The neutral soliton in
the same   theory, obtained by double analytic
continuation, was studied in Ref.~\cite{AlkacGuajardoOzsahin2025}. 

In three dimensions, the Maxwell potential grows logarithmically at infinity.
Changing the subtraction scale shifts its finite coefficient.  A finite
Maxwell boundary term changes the combination of coefficients held fixed.  The
dependence of charged AdS$_3$ conserved quantities on Maxwell boundary
conditions was treated in Ref.~\cite{PerezEtAl2015} and   Ref.~\cite{Clement2003}
discusses the arbitrary scale in the logarithmic gauge potential.
Ref.~\cite{FrassinoMannMureika2015} treats the thermodynamic role of the
mass renormalization length scale.  The AdS$_3$ Maxwell counterterm and the
relevant thermodynamic ensembles are discussed in Ref.~\cite{Jensen2011}.  We
keep the finite Maxwell boundary term explicit via the parameter $\cfin$
and specify the Maxwell boundary condition used to compare the actions.

For $\cfin=1$, we impose the grand-canonical boundary condition of
Ref.~\cite{Jensen2011}, which fixes the finite asymptotic coefficient $a_i$.
For other values of $\cfin$, we impose the mixed boundary condition defined
in Sec.~\ref{sec:action}.  Constant gauge potentials along the
noncontractible circles allow the black hole and soliton to satisfy the same
boundary data.
Their logarithmic coefficients point along different boundary directions.
Equality of both current components requires $q=Q=0$ for these two static
solutions.  We compare the charged static saddles under the stated boundary
condition.  We do not consider the decay of an isolated electrically charged
black hole.

\section{Electric black hole and magnetic soliton}
\label{sec:saddles}

We work with signature $(-,+,+)$ and set the gravitational constant $G_N=1$.  The Lorentzian
bulk action of the scalar--tensor formulation of three-dimensional
Einstein--Gauss--Bonnet gravity coupled to a Maxwell field is
\cite{HennigarKubiznakMann2021}
\begin{equation}
S_{\mathrm{bulk}}=\frac{1}{16\pi}\int_{\mathcal M}d^3x\sqrt{-g}
\left[R-2\Lambda+\alpha\mathcal L_{\phi}-F_{ab}F^{ab}\right],
\qquad
\Lambda=-\frac{1}{\ell^2}.
\label{eq:action}
\end{equation}
Here $F_{ab}=2\nabla_{[a}A_{b]}$, and
\begin{equation}
\mathcal L_{\phi}
=\phi\mathcal G+4G^{ab}\nabla_a\phi\nabla_b\phi
-4(\nabla\phi)^2\Box\phi
+2\bigl[(\nabla\phi)^2\bigr]^2 ,
\label{eq:Lphi}
\end{equation}
where $\mathcal G=R_{abcd}R^{abcd}-4R_{ab}R^{ab}+R^2$ and $G_{ab}$ is the
Einstein tensor.  In three spacetime
dimensions, $\mathcal G$ vanishes identically; technically the theory is a particular case of Horndeski theory, but it has come to be referred to as 3DEGB.  The derivative terms remain
in the action and are nonzero for the solutions considered below.  The
scalar does not add a propagating local degree of freedom in this theory
\cite{HennigarKubiznakMann2021}.  The three-dimensional action is invariant
under constant shifts of $\phi$.  We use this freedom to set the reference
length in the scalar profile to $\ell$.  The length $\ell$ is the bare AdS
scale defined by $\Lambda=-1/\ell^2$.  Equation~\eqref{eq:Balpha} defines the
effective asymptotic radius $\ella$.  The coupling $\alpha>0$ has dimensions
of length squared.
We shall use subscripts $b$ and $s$ to refer to black-hole and soliton quantities,
respectively.

After adding a flat connection along the spatial circle, we write the charged
Einstein-branch solution~\cite{HennigarKubiznakMann2021} as
\begin{align}
ds_b^2&=-f_b(r)dt_b^2+\frac{dr^2}{f_b(r)}+r^2d\psi_b^2,
&\phi&=\ln\frac r\ell ,
\nonumber\\
f_b+\frac{\alpha f_b^2}{r^2}
&=\frac{r^2}{\ell^2}-m_b-2q^2\ln\frac{r}{r_0},
&A_b&=-q\ln\frac r{r_+}\,dt_b+\chi_b\,d\psi_b 
\label{eq:bh}
\end{align}
where the Einstein-branch root is
\begin{equation}
f_b(r)=\frac{r^2}{2\alpha}
\left[
\sqrt{1+\frac{4\alpha}{r^2}
\left(
\frac{r^2}{\ell^2}-m_b-2q^2\ln\frac{r}{r_0}
\right)}
-1
\right],
\label{eq:bhbranch}
\end{equation}
which reduces continuously to the charged BTZ metric function as
$\alpha\to0$.  We choose the additive constant in $A_t$ so that
$A_t(r_+)=0$, ensuring regularity on the contractible Euclidean thermal
circle.  The scale $r_0$ specifies the subtraction convention
for the logarithmic solution.  The coordinate $\psi_b$ has period $\eta_b$,
$r_+$ is the largest root of $f_b$, and $\chi_b$ is a flat connection along
the spatial circle.
The horizon condition and thermodynamic quantities are
\begin{equation}
m_b=\frac{r_+^2}{\ell^2}+2q^2\ln\frac{r_0}{r_+},
\qquad
T=\frac{r_+^2-q^2\ell^2}{2\pi\ell^2r_+},
\qquad
S_b=\frac{\eta_b r_+}{4}.
\label{eq:bhthermo}
\end{equation}
A nonextremal black hole obeys $r_+>|q|\ell$.

As in Refs.~\cite{CataldoSalgado1996,AnabalonEtAl2022,AuzziEtAl2025}, the
analytic continuation acts on both the coordinates and the charge parameter:
\begin{equation}
t_b\to i\ell\psi_s,
\qquad
\psi_b\to \frac{i}{\ell}t_s,
\qquad
q\to iQ .
\label{eq:continuation}
\end{equation}
The continuation of $q$ makes the spatial gauge potential real and reverses
the sign of the logarithmic Maxwell term in the metric equation.  We set the
flat connection to zero for this continuation.  Writing the continued mass
parameter as $m_s$ and the zero of the metric function as $r_s$, we obtain
the real Lorentzian solution
\begin{align}
ds_s^2&=-\frac{r^2}{\ell^2}dt_s^2+\frac{dr^2}{f_s(r)}
+\ell^2f_s(r)d\psi_s^2,
&\phi&=\ln\frac r\ell ,
\nonumber\\
f_s+\frac{\alpha f_s^2}{r^2}
&=\frac{r^2}{\ell^2}-m_s+2Q^2\ln\frac{r}{r_0},
&A_s&=-\chi_s\,dt_s+\ell Q\ln\frac r{r_s}\,d\psi_s 
\label{eq:soliton}
\end{align}
where
\begin{equation}
f_s(r)=\frac{r^2}{2\alpha}
\left[
\sqrt{1+\frac{4\alpha}{r^2}
\left(
\frac{r^2}{\ell^2}-m_s+2Q^2\ln\frac{r}{r_0}
\right)}
-1
\right] .
\label{eq:solbranch}
\end{equation}
It is straightforward to check that   Eqs.~\eqref{eq:soliton} and \eqref{eq:solbranch} are solutions to
field equations of Eq.~\eqref{eq:action} for arbitrary $m_s$ and
$Q$.

A flat connection leaves the field equations unchanged, so independent real
constants $\chi_b$ and $\chi_s$ may be included when matching the boundary
data.  The parameter $Q$ is the coefficient of the spatial logarithmic
Maxwell mode,
\begin{equation}
F_{r\psi_s}=\frac{\ell Q}{r}.
\end{equation}
The corresponding flux through the soliton disk truncated at $r=\rho$ is
\begin{equation}
\mathcal F_s(\rho)
=\int_{r_s}^{\rho}dr\int_0^{\eta_s}d\psi_s\,F_{r\psi_s}
=\eta_s\ell Q\ln\frac{\rho}{r_s},
\label{eq:solitonfluxcutoff}
\end{equation}
which diverges as $\rho\to\infty$ when $Q\neq0$; an analogous divergence for
the Einstein--Maxwell magnetic soliton has been previously discussed ~\cite{AuzziEtAl2025}.  The parameter $Q$ does not represent a finite total
magnetic flux.

We take $r\geq r_s>0$, where the $\psi_s$ circle closes.  We choose the
additive constant in $A_{s,\psi}$ so that $A_{s,\psi}(r_s)=0$, as required
for a smooth one-form on this contractible circle.  The condition $f_s(r_s)=0$
fixes $m_s$, and differentiating the equation for $f_s$ fixes $f_s'(r_s)$.
Absence of a conical singularity at the cap then determines the period:
\begin{equation}
\begin{aligned}
m_s&=\frac{r_s^2}{\ell^2}-2Q^2\ln\frac{r_0}{r_s},
&
f_s'(r_s)&=\frac{2r_s}{\ell^2}+\frac{2Q^2}{r_s},\\
\eta_s&=\frac{4\pi}{\ell f_s'(r_s)}
=\frac{2\pi\ell r_s}{r_s^2+Q^2\ell^2}.
\end{aligned}
\label{eq:cap}
\end{equation}
This continuation constructs the soliton family.  The parameters of the two
real solutions are related only through the regularity conditions and the
matching of boundary data.

For both the black hole and soliton we have
\begin{equation}
f_b(r),f_s(r)=\frac{r^2}{\ella^2}+O(\ln r),
\end{equation}
where we define
\begin{equation}
\Ba=\frac{2}{1+\sqrt{1+4\alpha/\ell^2}},
\qquad
\ella=\frac{\ell}{\sqrt{\Ba}}
\label{eq:Balpha}
\end{equation}
and 
so $\ella$ is the effective asymptotic AdS radius.  

After continuing
$t_b=-i\tau_b$ and $t_s=-i\tau_s$, let $\beta_b$ and $\beta_s$ denote the
respective Euclidean periods.  Black-hole regularity fixes
$\beta_b=T^{-1}$.  The soliton period $\beta_s$ remains arbitrary until the
boundary tori are matched.
Using the common conformal factor $\ella^2/r^2$, the two Euclidean boundary
metrics are
\begin{equation}
d\bar s_b^2=d\tau_b^2+\ella^2d\psi_b^2,
\qquad
d\bar s_s^2=\frac{1}{\Ba}d\tau_s^2+\ell^2d\psi_s^2.
\label{eq:boundarymetrics}
\end{equation}
Matching the proper lengths of the corresponding thermal and spatial circles
gives
\begin{equation}
\beta_s=\sqrt{\Ba}\,\beta_b,
\qquad
\eta_s=\frac{\eta_b}{\sqrt{\Ba}},
\qquad
\beta_s\eta_s=\beta_b\eta_b .
\label{eq:periodmatch}
\end{equation}
Both fillings use $\phi=\ln(r/\ell)$ with the same additive constant, so their
scalar boundary data also agree.

\section{Maxwell boundary data and the Euclidean action}
\label{sec:action}

At a radial cutoff $r=\rho$, let $h_{ij}$ be the induced metric. 
Near the AdS boundary, the induced metric behaves as
\begin{equation}
h_{ij}
=
\frac{r^2}{\ella^2}
\left[
\gamma_{ij}^{(0)}
+O\!\left(
\frac{\ella^2\ln(r/r_0)}{r^2}
\right)
\right].
\end{equation}

Using the
conformal factor $\ella^2/\rho^2$, define the boundary metric in the chosen
conformal frame by
\begin{equation}
\gamma^{(0)}_{ij}
\equiv \lim_{\rho\to\infty}
\frac{\ella^2}{\rho^2}h_{ij}.
\label{eq:boundary-metric}
\end{equation}
For the black-hole and soliton fillings, Eq.~\eqref{eq:boundarymetrics} gives
the corresponding boundary metrics.

In radial gauge, an asymptotically AdS$_3$ Maxwell field has the expansion
\cite{Jensen2011}
\begin{equation}
A_i=\widetilde a_i\ln\frac{r}{r_0}+a_i+o(1).
\label{eq:falloff}
\end{equation}
Here $o(1)$ denotes terms that vanish as $r\to\infty$.  The indices $i,j$
label directions tangent to the surfaces of constant $r$.  Indices on the
asymptotic coefficients $a_i$ and $\widetilde a_i$ are raised with
$\gamma^{(0)ij}$, while indices on quantities at finite $r$ are raised with
$h^{ij}$.

The $r$ component of $\nabla_aF^{ab}=0$ is
$\partial_i(\sqrt{g}\,F^{ir})=0$.  Since
$h_{ij}=(r^2/\ella^2)\gamma^{(0)}_{ij}+o(r^2)$, Eqs.~\eqref{eq:boundary-metric}
and \eqref{eq:falloff} give
$\sqrt{g}\,F^{ir}
=-(\sqrt{\gamma^{(0)}}/\ella)\gamma^{(0)ij}\widetilde a_j+o(1)$.
The leading term therefore obeys
\begin{equation}
\partial_i\left(
\sqrt{\gamma^{(0)}}\,\gamma^{(0)ij}\widetilde a_j
\right)=0.
\label{eq:Maxwellconstraint}
\end{equation}

Eq.~\eqref{eq:Maxwellconstraint} constrains the logarithmic coefficient.
The boundary condition must be specified separately.  The conformally
invariant AdS$_3$ quantization fixes $\widetilde a_i$
\cite{MarolfRoss2006,Jensen2011}.  Jensen also defines a grand-canonical
ensemble in which the finite coefficient $a_i$ is fixed
\cite{Jensen2011}.  We use this grand-canonical boundary condition when
$\cfin=1$.  Here ``source'' denotes the datum fixed in the Euclidean
variational problem.

Define the boundary datum associated with the finite term by
\begin{equation}
a_i^{(\cfin)}=a_i+(1-\cfin)\widetilde a_i .
\label{eq:mixedsource}
\end{equation}
For $\cfin=1$, the fixed datum is $a_i$.  For $\cfin\neq1$, the variation
below gives the mixed condition $\delta a_i^{(\cfin)}=0$.
In the conformally invariant quantization, $\widetilde a_i$ is fixed.

At the same cutoff, let $n^a$ be the outward-pointing Euclidean unit normal
and define $\mathcal E_i=n^aF_{ai}$.  Indices on $\mathcal E_i$ are raised
with $h^{ij}$.  Adapting the logarithmic Maxwell counterterm of
Ref.~\cite{Jensen2011} to our normalization and adding a finite term, we use
\begin{equation}
I_{M,\mathrm{ct}}^{(\cfin)}
=-\frac{\ella}{8\pi}\int_{r=\rho}d^2x\sqrt h\,
\left[\ln\frac{\rho}{r_0}+\cfin-1\right]
\mathcal E_i\mathcal E^i .
\label{eq:Maxwellct}
\end{equation}
The logarithmic term cancels the divergence displayed below.  In Maxwell
variables, this counterterm contains a radial derivative and is not intrinsic
to the cutoff surface.  Jensen derives it by dualizing to a massless scalar,
for which the counterterm is local
\cite{Jensen2011}.  The finite term changes the boundary datum held fixed.
The Legendre transform to fixed current is given separately in
Eq.~\eqref{eq:currentLegendre} below.  We hold $r_0$, $\cfin$, and $\ella$ fixed
below.

We use $t=-i\tau$ and $I_E=-iS_L$, so the Maxwell part of the Euclidean bulk
action is
\begin{equation}
I_M=\frac{1}{16\pi}\int_{\mathcal M}d^3x\sqrt g\,F_{ab}F^{ab}.
\label{eq:EuclideanMaxwellaction}
\end{equation}
The Euclidean continuation of a real Lorentzian electric potential has an
imaginary temporal component.  For the Maxwell variation, $h_{ij}$ is held
fixed at the cutoff and $\gamma^{(0)}_{ij}$ at the boundary.  For either
filling,
$n^r=\rho/\ella+o(\rho)$ at large $\rho$.
In radial gauge, $F_{ri}=\partial_rA_i$.  Eq.~\eqref{eq:falloff}
therefore gives
\begin{equation}
\delta A_i
=\ln\frac{\rho}{r_0}\,\delta\widetilde a_i+\delta a_i+o(1),
\qquad
\sqrt h\,\mathcal E^i
=\frac{\sqrt{\gamma^{(0)}}}{\ella}\widetilde a^i+o(1),
\qquad
\delta\mathcal E_i
=\frac{1}{\ella}\delta\widetilde a_i+o(1).
\label{eq:Maxwellasymptoticvariation}
\end{equation}

On shell, the bulk Maxwell variation reduces to a boundary term.  The bulk
and counterterm variations are
\begin{equation}
\begin{aligned}
\left.\delta I_M\right|_{\mathrm{on\ shell}}
&=\frac{1}{4\pi}\int_{r=\rho}d^2x\sqrt h\,
\mathcal E^i\delta A_i\\
&=\frac{1}{4\pi\ella}\int d^2x\sqrt{\gamma^{(0)}}\,
\widetilde a^i
\left(
\delta a_i+\ln\frac{\rho}{r_0}\delta\widetilde a_i
\right)+o(1),\\
\delta I_{M,\mathrm{ct}}^{(\cfin)}
&=-\frac{\ella}{4\pi}\int_{r=\rho}d^2x\sqrt h\,
\left[\ln\frac{\rho}{r_0}+\cfin-1\right]
\mathcal E^i\delta\mathcal E_i\\
&=-\frac{1}{4\pi\ella}\int d^2x\sqrt{\gamma^{(0)}}\,
\widetilde a^i
\left(
\ln\frac{\rho}{r_0}+\cfin-1
\right)\delta\widetilde a_i+o(1).
\end{aligned}
\label{eq:Maxwellvariationpieces}
\end{equation}
In varying the counterterm, we used
$\delta(\mathcal E_i\mathcal E^i)
=2\mathcal E^i\delta\mathcal E_i$.

Adding these expressions cancels the terms proportional to
$\ln(\rho/r_0)$.  Removing the cutoff therefore gives the Maxwell
contribution to the renormalized variation,
\begin{equation}
\begin{aligned}
\delta I_{M,\mathrm{ren}}^{(\cfin)}
&=\frac{1}{4\pi\ella}\int d^2x\sqrt{\gamma^{(0)}}\,
\widetilde a^i
\left[
\delta a_i+(1-\cfin)\delta\widetilde a_i
\right]\\
&=\frac{1}{4\pi\ella}\int d^2x\sqrt{\gamma^{(0)}}\,
\widetilde a^i\delta a_i^{(\cfin)} .
\end{aligned}
\label{eq:variation}
\end{equation}
The second equality follows from Eq.~\eqref{eq:mixedsource}.  The Maxwell
variational principle is well posed when $\delta a_i^{(\cfin)}=0$, with
conjugate response $\widetilde a^i/(4\pi\ella)$.  The choice $\cfin=1$ fixes
$a_i$.  Other values fix the combination in Eq.~\eqref{eq:mixedsource} and
give mixed boundary conditions.
Appendix~\ref{app:radial} gives a radial derivation of the on-shell actions.

We match the sources after Euclidean continuation.  On the common boundary,
introduce the unit-period coordinates
$\widehat\tau=\tau_b/\beta_b=\tau_s/\beta_s$ and
$\widehat\psi=\psi_b/\eta_b=\psi_s/\eta_s$.  The gauge fields are
\begin{equation}
A_b^E=iq\ln\frac r{r_+}\,d\tau_b+\chi_b\,d\psi_b,
\qquad
A_s^E=i\chi_s\,d\tau_s+\ell Q\ln\frac r{r_s}\,d\psi_s.
\label{eq:Euclideangauge}
\end{equation}
Defining the unitless quantities
\begin{equation}
\Lb=\ln\frac{r_0}{r_+}-\cfin+1,
\qquad
\Ls=\ln\frac{r_0}{r_s}-\cfin+1 
\label{eq:logs}
\end{equation}
 the finite Maxwell sources are
\begin{equation}
\begin{aligned}
a_b^{(\cfin)}
 &=i\beta_bq\Lb\,d\widehat\tau
   +\eta_b\chi_b\,d\widehat\psi,\\
a_s^{(\cfin)}
 &=i\beta_s\chi_s\,d\widehat\tau
   +\eta_s\ell Q\Ls\,d\widehat\psi 
\end{aligned}
\label{eq:finitesources}
\end{equation}
pulled back to the common unit-period torus,
Equating the chosen representatives of the two boundary connections gives
\begin{equation}
\beta_b q\Lb=\beta_s\chi_s,
\qquad
\eta_b\chi_b=\eta_s\ell Q\Ls .
\label{eq:gaugematch}
\end{equation}
For compact $U(1)$, a large gauge transformation can shift the integral of a
real connection around a boundary circle without changing its Wilson loop.
Such a transformation is allowed only if it is smooth throughout the bulk
filling.  We make no such shift and impose Eq.~\eqref{eq:gaugematch} directly.
A real gauge transformation cannot change the imaginary temporal components
introduced by Wick rotation, so those components must agree exactly.

The terms proportional to $\chi_b$ and $\chi_s$ have zero field strength and
lie along the noncontractible circle of each solution: the spatial circle for
the black hole and the Euclidean thermal circle for the soliton.  They
represent flat connections with holonomy around these circles.  The
logarithmic terms vanish at the horizon or cap, where the corresponding
circle contracts.  The four-dimensional comparison~\cite{AnabalonEtAl2022} similarly includes a constant spatial component
of the black-hole gauge potential and a constant temporal component of the
soliton gauge potential.

Writing the full renormalized Euclidean action as
$I_{\mathrm{ren}}^{(\cfin)}$, we pass to the fixed-current, or Neumann,
ensemble by extending the $\cfin=1$ Legendre transform of
Ref.~\cite{Jensen2011} to the mixed datum in
Eq.~\eqref{eq:mixedsource}:
\begin{equation}
I_N^{(\cfin)}=I_{\mathrm{ren}}^{(\cfin)}
-\frac{1}{4\pi\ella}\int d^2x\sqrt{\gamma^{(0)}}\,
\widetilde a^i a_i^{(\cfin)} .
\label{eq:currentLegendre}
\end{equation}
At fixed $\gamma^{(0)}_{ij}$, its Maxwell variation is
\begin{equation}
\delta I_N^{(\cfin)}
=-\frac{1}{4\pi\ella}\int d^2x\sqrt{\gamma^{(0)}}\,
a_i^{(\cfin)}\delta\widetilde a^i .
\end{equation}
At $\cfin=1$, Eq.~\eqref{eq:currentLegendre} reduces to Jensen's renormalized
Legendre transform after translating conventions.  Jensen also shows that
performing the Legendre transform at a finite cutoff reverses the sign of the
logarithmic Maxwell counterterm.  For $\cfin\neq1$, the functional differs
from its $\cfin=1$ value by the finite term
\begin{equation}
I_N^{(\cfin)}-I_N^{(1)}
=\frac{\cfin-1}{8\pi\ella}\int d^2x\sqrt{\gamma^{(0)}}\,
\widetilde a_i\widetilde a^i .
\end{equation}
The current held fixed is
\begin{equation}
J^i=\frac{\widetilde a^i}{4\pi\ella}.
\label{eq:fixedcurrent}
\end{equation}
On the common unit-period torus, the logarithmic one-forms are
\begin{equation}
\widetilde a_b=i\beta_bq\,d\widehat\tau,
\qquad
\widetilde a_s=\eta_s\ell Q\,d\widehat\psi .
\label{eq:logmodes}
\end{equation}
The boundary metric $\gamma^{(0)}_{ij}$ is the same for the two fillings, so
equality of the currents is equivalent to equality of these one-forms.  Since
$d\widehat\tau$ and $d\widehat\psi$ are independent, equality of both current
components requires
\begin{equation}
q=Q=0 .
\label{eq:fixedcurrentneutrality}
\end{equation}
This conclusion applies to the two static solutions with both current
components fixed.  Other choices of fixed data or a larger family of
stationary solutions require a separate analysis.

Including the scalar--tensor Dirichlet term, the gravitational counterterm,
and Eq.~\eqref{eq:Maxwellct}, the renormalized actions are
\begin{align}
I_b^{(\cfin)}&=\frac{\beta_b\eta_b}{16\pi}
\left[-\frac{r_+^2}{\ell^2}+2q^2(1-\Lb)\right],
\nonumber\\
I_s^{(\cfin)}&=\frac{\beta_s\eta_s}{16\pi}
\left[-\frac{r_s^2}{\ell^2}-2Q^2(1-\Ls)\right].
\label{eq:individualactions}
\end{align}
Their difference is
\begin{equation}
\Delta I^{(\cfin)}\equiv I_b^{(\cfin)}-I_s^{(\cfin)}
=\frac{\beta_b\eta_b}{16\pi}
\left[
\frac{r_s^2-r_+^2}{\ell^2}
+2q^2(1-\Lb)
+2Q^2(1-\Ls)
\right]
\label{eq:DeltaI}
\end{equation}
where \eqref{eq:periodmatch} has been used.
Both charge terms enter $\Delta I^{(\cfin)}$ with plus signs, multiplied by $1-\Lb$ and $1-\Ls$, respectively.  These factors determine the numerical signs of the two terms.

 {In the boundary metrics~\eqref{eq:boundarymetrics}, the thermal circle
has proper length $\beta_b$ for the black hole and $\beta_s/\sqrt{\Ba}$ for the soliton, and Eq.~\eqref{eq:periodmatch} sets these equal.  Both saddles are therefore at the temperature $T=\beta_b^{-1}$, and we define their
thermodynamic potentials with this common period,}
\begin{equation}\label{eq:Gdef}
G_b^{(\cfin)}=\frac{I_b^{(\cfin)}}{\beta_b},
\qquad
G_s^{(\cfin)}=\frac{I_s^{(\cfin)}}{\beta_b}
\end{equation}
{The difference $G_b^{(\cfin)}-G_s^{(\cfin)}=\Delta I^{(\cfin)}/\beta_b$
then has the sign of $\Delta I^{(\cfin)}$.  The ratio
$I_s^{(\cfin)}/\beta_s=G_s^{(\cfin)}/\sqrt{\Ba}$ is instead the potential
associated with the coordinate time $\tau_s$, whose normalization is set by
the continuation~\eqref{eq:continuation}.  Comparing it with
$G_b^{(\cfin)}$ would mix two normalizations of time unless $\alpha=0$.}

Equations~\eqref{eq:bhthermo} and
\eqref{eq:cap} then give
\begin{align}
G_b^{(\cfin)}
&=\frac{\eta_b}{16\pi}\left[-m_b+2\cfin q^2\right],
&
G_s^{(\cfin)}
&=\frac{\eta_b}{16\pi}\left[-m_s-2\cfin Q^2\right],
\label{eq:individualmassform}
\end{align}
and hence
\begin{equation}
\Delta G^{(\cfin)}
=\frac{\eta_b}{16\pi}
\left[m_s-m_b+2\cfin(q^2+Q^2)\right].
\label{eq:massform}
\end{equation}
At $\cfin=0$, Eq.~\eqref{eq:massform} gives
\begin{equation}
\Delta G^{(0)}=\frac{\eta_b}{16\pi}(m_s-m_b).
\end{equation}
The dependence on the black-hole and soliton mass parameters is then the same,
up to notation and overall normalization, as Eq.~(3.5) in the
published version of Ref.~\cite{AnabalonEtAl2022}.  This is a comparison of
the formulas; the boundary conditions are different.  Charge dependence
remains implicit through the regularity and fixed-source equations.

The black-hole thermodynamic energy derived from $G_b^{(\cfin)}$ is
\begin{equation}
E_b^{(\cfin)}
=\frac{\eta_b}{16\pi}
\left[m_b-2(\cfin-1)q^2\right]
=\frac{\eta_b}{16\pi}
\left[\frac{r_+^2}{\ell^2}+2q^2\Lb\right].
\label{eq:bhenergy}
\end{equation}
At fixed $(\ell,\alpha,r_0,\cfin,\eta_b)$, define the electric charge and its
conjugate potential by
\begin{equation}
q_b=\frac{\eta_bq}{4\pi},
\qquad
\mu_b=q\Lb .
\label{eq:chargepotential}
\end{equation}
Direct differentiation gives
\begin{equation}
dE_b^{(\cfin)}=T\,dS_b+\mu_b\,dq_b,
\qquad
G_b^{(\cfin)}
=E_b^{(\cfin)}-TS_b-\mu_bq_b .
\label{eq:firstlawcheck}
\end{equation}
For $\eta_b=2\pi$, $q_b=q/2$.  At $\cfin=1$,
$E_b=m_b/8$ and $S_b=\pi r_+/2$, in agreement with
Ref.~\cite{HennigarKubiznakMann2021}.  In our normalization,
\begin{equation}
(q_b,\mu_b)
=\left(\frac q2,q\ln\frac{r_0}{r_+}\right).
\end{equation}
In Ref.~\cite{HennigarKubiznakMann2021}, the physical charge has numerical
value $q$ in our parametrization, and its conjugate potential is
$q\ln(r_0/r_+)/2$.  Both conventions give the same work term in the first
law.

For any constant $\kappa$, a change in the logarithmic reference scale can be
compensated by
\begin{equation}
r_0\longrightarrow r_0e^\kappa,
\qquad
a_i\longrightarrow a_i+\kappa\widetilde a_i,
\qquad
\cfin\longrightarrow\cfin+\kappa,
\qquad
m_b\longrightarrow m_b+2q^2\kappa,
\qquad
m_s\longrightarrow m_s-2Q^2\kappa .
\label{eq:schemetransform}
\end{equation}
Together, these transformations leave the geometries, fixed sources,
Eq.~\eqref{eq:bhenergy}, and
Eqs.~\eqref{eq:individualactions}--\eqref{eq:massform} unchanged.  The shift
in $a_i$ accounts for the change in the finite coefficient when the logarithm
is written with the scale $r_0e^\kappa$.  These transformations amount to a
reparametrization.  Changing $r_0$ alone changes the fixed source; at fixed
mass parameters, it also changes the charged solutions.

\section{Fixed-source thermodynamics}
\label{sec:phase}

Set $\eta_b=2\pi$ and introduce
\begin{equation}
x=\frac{r_+}{\ell\sqrt{\Ba}},
\quad
y=\frac{r_s}{\ell\sqrt{\Ba}},
\quad
\bar r_0=\frac{r_0}{\ell\sqrt{\Ba}} .
\label{eq:dimensionless}
\end{equation}
Then
\begin{equation}
\Lb=\ln\frac{\bar r_0}{x}-\cfin+1,
\qquad
\Ls=\ln\frac{\bar r_0}{y}-\cfin+1.
\label{eq:dimensionlesslogs}
\end{equation}
In this section we suppress the superscript $(\cfin)$ on $G_b$, $G_s$, and
their rescaled forms.
Let $\mu_b=q\Lb$ denote the electric potential conjugate to
$q_b$.  We define the reduced sources by
\begin{equation}
\varphi=\frac{\mu_b}{2\sqrt{\Ba}}
=\frac{\chi_s}{2}
=\frac{q\Lb}{2\sqrt{\Ba}},
\qquad
\Phi=\frac{\chi_b}{\ell}=\frac{Q\Ls}{\sqrt{\Ba}}.
\label{eq:sources}
\end{equation}
Eq.~\eqref{eq:sources} gives $\mu_b=2\sqrt{\Ba}\,\varphi$.  The quantity
$\Phi$ is the dimensionless coefficient of the flat connection along the
spatial circle.  For $\eta_b=2\pi$, the corresponding line integral is
\begin{equation}
\oint_{\psi_b}A_b=\eta_b\chi_b=2\pi\ell\Phi .
\end{equation}
For $\Lb\Ls\neq0$, the fixed-source equations can be written 
\begin{equation}
\frac{q}{\sqrt{\Ba}}=\frac{2\varphi}{\Lb},
\qquad
\frac{Q}{\sqrt{\Ba}}=\frac{\Phi}{\Ls},
\qquad
\widehat T\equiv2\pi\ella T=x-\frac{q^2}{\Ba x},
\qquad
y^2-y+\frac{Q^2}{\Ba}=0
\label{eq:fixedsourceeqs}
\end{equation}
uwing Eqs.~\eqref{eq:sources}, \eqref{eq:bhthermo}, and \eqref{eq:cap}
with \eqref{eq:periodmatch}.
The last equation is
the cap condition~\eqref{eq:cap} with $\eta_s=\eta_b/\sqrt{\Ba}=2\pi/\sqrt{\Ba}$
from Eq.~\eqref{eq:periodmatch}.  At fixed $(\varphi,\Phi)$, the black-hole
relations give $\widehat T$ as a function of $x$, while the soliton relations
determine $y$ and $Q$ independently of the temperature.

If $\varphi=0$ or $\Phi=0$, the cases with the corresponding $\Lb=0$ or
$\Ls=0$ must be obtained directly from Eq.~\eqref{eq:sources}, because
Eq.~\eqref{eq:fixedsourceeqs} divides by these quantities.
The rescaled thermodynamic potentials
\begin{equation}
\widetilde G_b=\frac{8}{\Ba}G_b,
\qquad
\widetilde G_s=\frac{8}{\Ba}G_s.
\label{eq:scaleddefinition}
\end{equation}
take the form
\begin{equation}
\widetilde G_b=-x^2+\frac{2q^2}{\Ba}(1-\Lb),
\qquad
\widetilde G_s=-y^2-\frac{2Q^2}{\Ba}(1-\Ls).
\label{eq:scaledG}
\end{equation}

As a check, on the neutral branch $q=Q=0$, cap regularity gives $y=1$.
Equality of the two thermodynamic potentials then gives $x=1$ and
\begin{equation}
\widehat T_c=1,
\qquad
T_c=\frac{1}{2\pi\ella}.
\end{equation}
This is the EGB counterpart of the neutral black-hole--AdS-soliton transition
studied in Ref.~\cite{SuryaSchleichWitt2001}.  The neutral soliton
normalization agrees with Ref.~\cite{AlkacGuajardoOzsahin2025}. The limit $\alpha\to0$ is discussed in Sec.~\ref{sec:alpha0}.

The fixed-source equations can have several roots.  At fixed
$(\varphi,\Phi,\bar r_0,\cfin)$, the dimensionless curves are independent of
$\Ba$.  The dependence on $\alpha$ enters through the conversion to physical
temperatures, sources, lengths, and thermodynamic potentials.  We restrict the
analysis to nonextremal black holes, $x>|q|/\sqrt{\Ba}$, and smooth solitons,
$Q^2/\Ba\leq1/4$.  Nonzero sources also require $\Lb\neq0$ and $\Ls\neq0$.

For the black hole we impose positive heat capacity at fixed $\varphi$ and
positive isothermal electric susceptibility.  In the variables above these
conditions are
\begin{equation}
x^2+\frac{q^2}{\Ba}\left(1-\frac{2}{\Lb}\right)>0,
\qquad
\Lb-\frac{2q^2}{\Ba x^2+q^2}>0.
\label{eq:BHresponse}
\end{equation}
For each soliton root, we require
$d(Q/\sqrt{\Ba})/d\Phi>0$.  Away from a turning point, this is equivalent to
\begin{equation}
\frac{d\Phi}{d(Q/\sqrt{\Ba})}
=\Ls+\frac{2(1-y)}{2y-1}>0.
\label{eq:solitonresponse}
\end{equation}
The double root $y=1/2$ is treated only as a limit.  These response conditions
apply within the static ansatz.  A full stability analysis requires a
calculation of the fluctuation spectrum.

\begin{figure}[!ht]
\centering
\includegraphics[width=\textwidth]{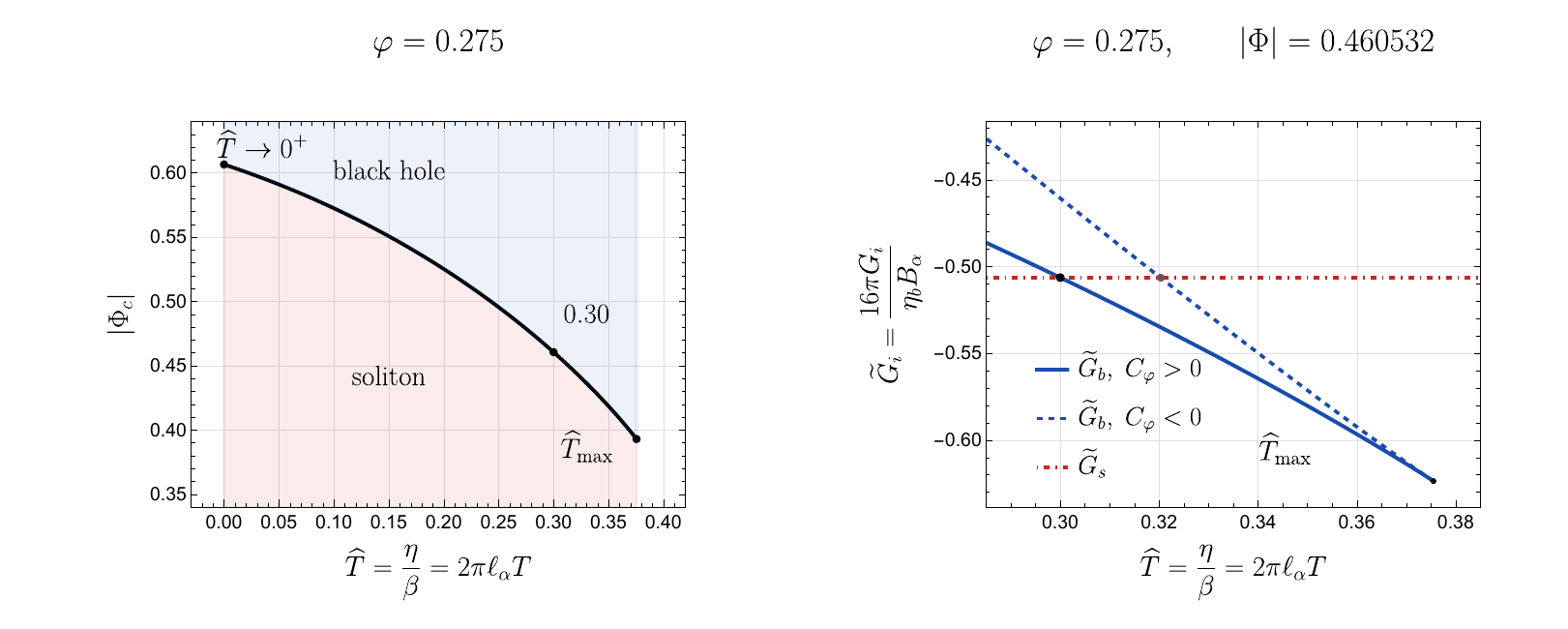}
\caption{Coexistence curve and black-hole branches for $\cfin=1$,
$\bar r_0=2$, and $\varphi=0.275$.  In the horizontal-axis label,
$\beta\equiv\beta_b=\beta_s/\sqrt{\Ba}=T^{-1}$ and
$\eta\equiv\ella\eta_b=\ell\eta_s=2\pi\ella$ are the common proper lengths
of the thermal and spatial boundary circles.  Left: the value $|\Phi_c|$ at which the
black-hole branch satisfying Eq.~\eqref{eq:BHresponse} and the soliton with
the lowest $\widetilde G_s$ satisfying Eq.~\eqref{eq:solitonresponse} have
equal thermodynamic potentials.  Among these two static saddles, the
soliton has lower thermodynamic potential below the curve and the black hole
above it.  The black points mark the limiting $\widehat T\to0^+$ endpoint,
coexistence at the chosen temperature $\widehat T=0.30$, and the limiting
fold endpoint $\widehat T\to\widehat T_{\max}^{-}$.
Right: the thermodynamic potentials at fixed $|\Phi|\simeq0.460532$.  The
solid and dashed blue curves are the black-hole
branches with $C_\varphi>0$ and $C_\varphi<0$, respectively, and the red
dot-dashed line is the soliton.  The black point on the soliton line is the
same $\widehat T=0.30$ coexistence point marked in the left panel.  The
gray point is a different equality: at the same fixed $|\Phi|$, the
$C_\varphi<0$ branch meets the soliton at
$\widehat T\simeq0.320268$.  The gray marker is used because this equality is
excluded from the phase boundary shown in the left panel.  It is not a phase
transition because this branch fails Eq.~\eqref{eq:BHresponse} and the
$C_\varphi>0$ black hole has lower thermodynamic potential at the same
temperature.  At the rightmost black point, the two black-hole branches meet
and both inequalities in Eq.~\eqref{eq:BHresponse} are saturated.}
\label{fig:phaseboundary}
\end{figure}

The soliton has no horizon. For fixed boundary sources, its action is
proportional to the thermal period, which gives $S_s=0$.  With
$(\ell,\alpha,r_0,\cfin,\eta_b)$ held fixed, fixing $(\varphi,\Phi)$ is
equivalent to fixing $(\mu_b,\chi_b)$ by Eq.~\eqref{eq:sources}.  Therefore
\begin{equation}
\left.
\frac{\partial(G_b-G_s)}{\partial T}
\right|_{\mu_b,\chi_b}
=-(S_b-S_s)=-S_b .
\label{eq:firstorder}
\end{equation}
At the black coexistence point shown in the right panel of
Fig.~\ref{fig:phaseboundary}, $S_b>0$, so the two curves have different
slopes.  The transition between these two saddles is first order, with
entropy change $S_b$ and latent heat $T S_b$, evaluated at the crossing.
Other smooth saddles and the fluctuation spectrum must also be checked.

\subsection{Finite-temperature coexistence}
\label{subsec:finitecoexistence}

We choose $\widehat T=0.30$, or $T=0.30/(2\pi\ella)$, as a representative
finite temperature; the   phase behavior is qualitatively the same as
the results we obtain for other values of $\widehat T \neq 0$.  The corresponding coexistence point is marked in both
panels of Fig.~\ref{fig:phaseboundary}.
Equating the free energies of the black-hole branch satisfying
Eq.~\eqref{eq:BHresponse} and the lowest-free-energy soliton satisfying
Eq.~\eqref{eq:solitonresponse} gives
\begin{equation}
\begin{aligned}
x&\simeq0.684315,
&y&\simeq0.718387,
&\Phi_c&\simeq\pm0.460532,\\
\widetilde G_b&=\widetilde G_s\simeq-0.506413.
\end{aligned}
\label{eq:T030coexistence}
\end{equation}
The two signs follow from the symmetry $(Q,\Phi)\to(-Q,-\Phi)$.  The other
soliton satisfying Eq.~\eqref{eq:solitonresponse} has higher free energy.

Keeping $|\Phi|\simeq0.460532$ fixed at the value in
Eq.~\eqref{eq:T030coexistence}, a separate equality occurs on the
$C_{\varphi}<0$ branch at $\widehat T\simeq0.320268$, as shown by the gray
point in the right panel of Fig.~\ref{fig:phaseboundary}
where the dashed and dot-dashed curves intersect
.  This is about
$2.03\times10^{-2}$ above the chosen coexistence temperature.  The
$C_{\varphi}>0$ black hole has lower free energy there, so this second
equality is not a phase transition and is not part of the coexistence curve
in the left panel.
At $\widehat T=0.30$, the second black-hole root occurs at
$x\simeq1.140252$, with $\widetilde G_b\simeq-0.460685$.  It has
$C_{\varphi}<0$ and fails both conditions in Eq.~\eqref{eq:BHresponse}.  If
the spatial source is varied at this temperature, its free energy equals that
of the soliton at $\Phi\simeq\pm0.485657$, but the
$C_{\varphi}>0$ black hole again has lower free energy.  In the limit
$\widehat T\to\widehat T_{\max}^{-}$, the coexistence curve in the left
panel approaches
\begin{equation}
\widehat T_{\max}\simeq0.375498,
\qquad
\Phi_c\simeq\pm0.392950.
\label{eq:foldcoexistence}
\end{equation}
Here $\Phi$ is the dimensionless spatial Maxwell source.  The line integral
around the black-hole spatial circle is $2\pi\ell\Phi$.  This is not the total
magnetic flux through the soliton disk, which diverges
logarithmically for $Q\neq0$ by Eq.~\eqref{eq:solitonfluxcutoff}.

\subsection{Black-hole branches and the maximum temperature}
\label{subsec:fold}

The maximum temperature displayed in both panels of
Fig.~\ref{fig:phaseboundary} follows from fixing the electric potential.
Since $\Lb$ decreases as $x$ increases, keeping
$\varphi=q\Lb/(2\sqrt{\Ba})$ fixed requires the magnitude of $q$ to increase.
The charge contribution eventually reverses the increase of temperature.  At
fixed $(\varphi,\bar r_0,\cfin)$,
\begin{equation}
\widehat T(x)=x-\frac{4\varphi^2}{x\Lb^2},
\qquad
\frac{d\widehat T}{dx}
=1+\frac{4\varphi^2}{x^2\Lb^2}
\left(1-\frac{2}{\Lb}\right).
\label{eq:temperaturederivative}
\end{equation}
This effect is absent at fixed $q$, for which
$(\partial\widehat T/\partial x)_q=1+q^2/(\Ba x^2)>0$.  Since
$S_b=\pi\ell\sqrt{\Ba}\,x/2$, the fixed-potential heat capacity is
\begin{equation}
C_{\varphi}
\equiv T\left(\frac{\partial S_b}{\partial T}\right)_{\varphi}
=\frac{\pi\ell\sqrt{\Ba}}{2}
\frac{\widehat T}{d\widehat T/dx},
\label{eq:heatcapacity}
\end{equation}
where the other boundary data are held fixed.  The two expressions on the
left-hand side of Eq.~\eqref{eq:BHresponse} can be written as
\begin{equation}
\begin{aligned}
x^2+\frac{q^2}{\Ba}\left(1-\frac{2}{\Lb}\right)
&=x^2\frac{d\widehat T}{dx},\\
\Lb-\frac{2q^2}{\Ba x^2+q^2}
&=\frac{\Lb x^2}{x^2+q^2/\Ba}
\frac{d\widehat T}{dx}.
\end{aligned}
\label{eq:responseslope}
\end{equation}
Thus, on the plotted $\Lb>0$ branch, both left-hand sides of
Eq.~\eqref{eq:BHresponse} vanish when $d\widehat T/dx=0$.  The turning point
$x_*$ obeys
\begin{equation}
x_*^2 \Lb(x_*)^3
=4\varphi^2\left[2-\Lb(x_*)\right],
\qquad
\Lb(x_*)=\ln\frac{\bar r_0}{x_*}-\cfin+1.
\label{eq:foldcondition}
\end{equation}
For $\cfin=1$, $\bar r_0=2$, and $\varphi=0.275$, this gives
\begin{equation}
x_*\simeq0.943308,
\qquad
\widehat T_{\max}=\widehat T(x_*)\simeq0.375498,
\qquad
T_{\max}=\frac{\widehat T_{\max}}{2\pi\ella}.
\label{eq:foldvalue}
\end{equation}
At this point $d^2\widehat T/dx^2\simeq-2.96079<0$.  For each
$0<\widehat T<\widehat T_{\max}$, the equation
$\widehat T=\widehat T(x)$ has two roots with $\Lb>0$.  The root with $x<x_*$
satisfies Eq.~\eqref{eq:BHresponse}.  The root with $x>x_*$
has $C_{\varphi}<0$, fails the electric response condition, and reaches
$\widehat T=0$ at $x\simeq1.317396$.  There is no $\Lb>0$ root above
$\widehat T_{\max}$.  A separate $\Lb<0$ branch, separated by the pole at
$\Lb=0$, also fails the electric susceptibility condition.  Thus
Eq.~\eqref{eq:foldvalue} is not a universal upper bound on black-hole
temperature.

For fixed sources,
\begin{equation}
\begin{aligned}
\left.
\frac{\partial^2G_b}{\partial T^2}
\right|_{\varphi,\Phi}
&=-\frac{C_{\varphi}}{T},
&
\frac{d\widetilde G_b}{d\widehat T}&=-2x,\\
\frac{d^2\widetilde G_b}{d\widehat T^2}
&=-\frac{2}{d\widehat T/dx}
=-\frac{4C_{\varphi}}
{\pi\ell\sqrt{\Ba}\,\widehat T}.
\end{aligned}
\label{eq:freeenergycurvature}
\end{equation}
As shown in the right panel of Fig.~\ref{fig:phaseboundary}, both black-hole branches
have negative slope, but their curvatures have opposite signs.  
The branch with $C_{\varphi}>0$ is concave downward, while the branch with
$C_{\varphi}<0$ is concave upward.  Their curvatures tend to $-\infty$ and
$+\infty$, respectively, at $x=x_*$, where both slopes approach $-2x_*$.  The
maximum-temperature point is therefore not an inflection point.  At
$\widehat T=0.30$, the negative-heat-capacity solution has a higher rescaled
thermodynamic potential than the positive-heat-capacity solution, with
$\Delta\widetilde G_b\simeq4.57\times10^{-2}$.  More generally, integrating
$d\widetilde G_b/d\widehat T=-2x$ from $\widehat T_{\max}$ shows that the
solution with $x>x_*$ has higher free energy for every common temperature
$0<\widehat T<\widehat T_{\max}$, with equality only at the fold.
Its crossing with the soliton therefore does not produce a phase transition.

\subsection{Zero-temperature coexistence endpoint}

The $\widehat T\to0^+$ endpoint in the left panel of
Fig.~\ref{fig:phaseboundary} occurs at two opposite values of the spatial
source $\Phi$.  We find these values by taking the $T\to0^+$ limit of
$G_b=I_b/\beta_b$.  Taking the nonextremal black-hole branch to extremality,
Eq.~\eqref{eq:fixedsourceeqs} gives
\begin{equation}
x_{\mathrm{ext}}^2=\frac{q^2}{\Ba},
\qquad
x_{\mathrm{ext}}^2\Lb(x_{\mathrm{ext}})^2=4\varphi^2,
\qquad
\widetilde G_b(0^+)\equiv\lim_{T\to0^+}\widetilde G_b
=x_{\mathrm{ext}}^2\left[1-2\Lb(x_{\mathrm{ext}})\right].
\label{eq:extremalendpoint}
\end{equation}
The soliton remains smooth and its thermodynamic potential is independent of
temperature at fixed $\Phi$.  Its cap condition and spatial source give
\begin{equation}
\frac{Q^2}{\Ba}=y(1-y),
\qquad
\Phi=\frac{Q}{\sqrt{\Ba}}\Ls .
\label{eq:zerosoliton}
\end{equation}
Consequently, the $T\to0^+$ coexistence limit requires
\begin{equation}
x_{\mathrm{ext}}^2\left[1-2\Lb(x_{\mathrm{ext}})\right]
=-y^2-2y(1-y)(1-\Ls).
\label{eq:zerotemperaturecoexistence}
\end{equation}
For $\cfin=1$, $\bar r_0=2$, $\varphi=0.275$, and $\Phi>0$, the
$T\to0^+$ black-hole limit and the soliton with the lowest $\widetilde G_s$ satisfying
Eq.~\eqref{eq:solitonresponse} obey these equations at
\begin{equation}
\begin{aligned}
x_{\mathrm{ext}}&\simeq0.279471,
&y&\simeq0.584240,
&\Phi_0&\simeq0.606500,\\
\frac{Q}{\sqrt{\Ba}}&\simeq0.492853,
&\widetilde G_b(0^+)&=\widetilde G_s\simeq-0.229314.
\end{aligned}
\label{eq:zeroexample}
\end{equation}
Here $Q^2/\Ba\simeq0.242904<1/4$, and the soliton satisfies
Eq.~\eqref{eq:solitonresponse}.  Moreover,
\begin{equation}
\left.
\frac{\partial(\widetilde G_b-\widetilde G_s)}{\partial\Phi}
\right|_{T\to0^+,\varphi}
=-\frac{4Q}{\sqrt{\Ba}}\neq0,
\label{eq:zerosourceslope}
\end{equation}
at $\Phi=\Phi_0$.  The symmetry $(Q,\Phi)\to(-Q,-\Phi)$ gives the second
endpoint, so $\Phi\simeq\pm0.606500$.  For $\Phi_0>0$,
Eq.~\eqref{eq:zerosourceslope} is negative.  Among these two saddles, the
soliton has lower thermodynamic potential just below $\Phi_0$, and the black
hole just above it.

Since physical temperatures satisfy $T\geq0$, the crossing is reached by
varying $\Phi$.  At fixed $(\varphi,\Phi)=(0.275,\Phi_0)$,
Eq.~\eqref{eq:firstorder} gives
\begin{equation}
G_b-G_s=-S_{\mathrm{ext}}T+O(T^2),
\qquad
S_{\mathrm{ext}}
=\frac{\pi\ell\sqrt{\Ba}}{2}x_{\mathrm{ext}}>0,
\label{eq:zerotemperatureexpansion}
\end{equation}
so, among these two saddles, the black hole has lower thermodynamic potential
immediately above zero temperature.  Along the coexistence line, the latent
heat tends to zero, while $S_b-S_s$ tends to the nonzero value
$S_{\mathrm{ext}}$.

Within this $T\to0^+$ two-saddle comparison, the first derivative of the
thermodynamic potential with respect to $\Phi$ is discontinuous at $\Phi_0$,
which is the usual first-order criterion.  Here $S_{\mathrm{ext}}$ is the entropy approached
along the nonextremal branch.  An exactly extremal Euclidean black hole must
be treated separately because regularity does not fix the period of its
thermal circle \cite{HawkingHorowitzRoss1995}.  Other saddles and their
fluctuation spectra are needed to establish the full phase diagram.

\section{The limit \texorpdfstring{$\alpha\to0$}{alpha -> 0}}
\label{sec:alpha0}

 As $\alpha\to0$, Eq.~\eqref{eq:Balpha} gives $\Ba\to1$ and
$\ella\to\ell$, and Eqs.~\eqref{eq:bhbranch} and \eqref{eq:solbranch} reduce
to
\begin{equation}
f_b=\frac{r^2}{\ell^2}-m_b-2q^2\ln\frac{r}{r_0},
\qquad
f_s=\frac{r^2}{\ell^2}-m_s+2Q^2\ln\frac{r}{r_0}
\label{eq:alpha0}
\end{equation}
which are the charged BTZ black hole~\cite{BanadosTeitelboimZanelli1992} and
the static magnetic Einstein--Maxwell solution~\cite{CataldoSalgado1996,AuzziEtAl2025}.  The scalar appears
in Eq.~\eqref{eq:action} only through $\alpha\mathcal L_\phi$ and drops out at
$\alpha=0$.  Eq.~\eqref{eq:periodmatch} becomes $\beta_s=\beta_b$ and
$\eta_s=\eta_b$, so $\tau_s$ is already the proper time of the boundary metric
and $I_s^{(\cfin)}/\beta_s=G_s^{(\cfin)}$.

{The actions~\eqref{eq:individualactions} contain no explicit $\alpha$,
and the reduced relations~\eqref{eq:fixedsourceeqs}, \eqref{eq:scaledG},
\eqref{eq:BHresponse}, and \eqref{eq:solitonresponse} contain it only through
$q/\sqrt{\Ba}$ and $Q/\sqrt{\Ba}$.  As noted in Sec.~\ref{sec:phase}, the
reduced curves at fixed $(\varphi,\Phi,\bar r_0,\cfin)$ are therefore the
same for every $\alpha\geq0$.  Fig.~\ref{fig:phaseboundary} also applies to
Einstein--Maxwell gravity, with $\Ba=1$ and $\ella=\ell$.}

{For $q=Q=0$ and $m_b=m_s$, the coordinate transformation
\begin{equation}
\tau_b=\ell\psi_s,
\qquad
\psi_b=\frac{\tau_s}{\ell}
\label{eq:exchange}
\end{equation}
maps the Euclidean black hole, including its scalar, onto the Euclidean
soliton for every $\alpha$.  It exchanges the thermal and spatial boundary
circles.  Denoting their proper lengths by $\beta\equiv\beta_b$ and
$\eta\equiv\ella\eta_b$, as in Fig.~\ref{fig:phaseboundary},
equations~\eqref{eq:bhthermo}, \eqref{eq:cap}, \eqref{eq:periodmatch}, and
\eqref{eq:individualactions} then give
\begin{equation}
I_b=-\frac{\pi\Ba\ella}{4}\,\frac{\eta}{\beta},
\qquad
I_s=-\frac{\pi\Ba\ella}{4}\,\frac{\beta}{\eta}.
\label{eq:neutralactions}
\end{equation}
Exchanging the two circles interchanges these actions, which are equal at
$\beta=\eta$.  For $\eta_b=2\pi$ this is $\widehat T=1$, the neutral
transition found in Sec.~\ref{sec:phase}.  At $\alpha=0$ the neutral soliton
is global AdS$_3$~\cite{SuryaSchleichWitt2001}, and
Eq.~\eqref{eq:neutralactions} with $\eta_b=2\pi$ gives
$I_b=-\pi^2\ell^2/(2\beta)$ and $I_s=-\beta/8$.  The transition is then the
Hawking--Page transition between the BTZ black hole and thermal AdS$_3$ at
$T=1/(2\pi\ell)$~\cite{MaldacenaStrominger1998,EuneKimYi2013}.  The
transformation~\eqref{eq:exchange} relates saddles with exchanged boundary
data, not the two saddles at fixed $(\beta,\eta)$, so it fixes the location
of the neutral transition.  For $q\neq0$, it maps the electric black hole to a
soliton with imaginary $Q$.  The magnetic soliton is obtained only after
$q\to iQ$, and the charged coexistence curve is fixed by the regularity and
source conditions of Sec.~\ref{sec:phase} rather than by a symmetry.}

\section{Discussion}

At $\cfin=1$, Eq.~\eqref{eq:massform} implies
\begin{equation}
G_b<G_s
\quad\Longleftrightarrow\quad
m_b>m_s+2(q^2+Q^2).
\label{eq:dominance}
\end{equation}
At fixed $m_b$ and $m_s$, increasing either $q^2$ or $Q^2$ raises $G_b-G_s$
and disfavors the black-hole saddle.  In this algebraic comparison, the
boundary sources vary.  Along a fixed-source branch in the grand-canonical
ensemble, $q$, $Q$, $r_+$, $r_s$, $m_b$, and $m_s$ vary together.  At
$\cfin=0$,
Eq.~\eqref{eq:massform} contains no explicit charge terms.  Charge dependence
remains through the regularity and source equations.

For $Q\neq0$, 
a change in the logarithmic reference scale 
changes $m_s$ via Eq.~\eqref{eq:schemetransform}. The geometry,
source, and renormalized action are invariant under this transformation, so
$m_s$ is scheme dependent.  Because the soliton logarithmic one-form in
Eq.~\eqref{eq:logmodes} has no temporal component, the temporal Maxwell source
has no conjugate soliton charge.  On each smooth soliton branch, cap regularity
and the fixed spatial source determine $r_s$ and $Q$ independently of
$\beta_b$.  Since
$I_s^{(\cfin)}$ is proportional to $\beta_b$,
\begin{equation}
S_s=0,
\qquad
E_s^{(\cfin)}
=\left.
\frac{\partial I_s^{(\cfin)}}{\partial\beta_b}
\right|_{\mu_b,\chi_b,\eta_b,\ell,\alpha,r_0,\cfin}
=G_s^{(\cfin)}
=-\frac{\eta_b}{16\pi}\left(m_s+2\cfin Q^2\right).
\label{eq:solitonenergy}
\end{equation}
This energy is conjugate to the common boundary time at fixed spatial Maxwell
source.  Relating it to a conserved mass requires a separate calculation using
the renormalized boundary stress tensor, with $a_i^{(\cfin)}$ held fixed during
the metric variation.
We do not perform that calculation here.  The combination in
Eq.~\eqref{eq:solitonenergy} is invariant under
Eq.~\eqref{eq:schemetransform}.  In the neutral limit with $\eta_b=2\pi$,
regularity gives $m_s=\Ba$ and $E_s=-\Ba/8$.  This agrees with Eq.~(2.43) of
Ref.~\cite{AlkacGuajardoOzsahin2025} after identifying $f_\infty$ in that
reference with $\Ba$ and setting $G_N=1$.

When parametrized by $(r_+,r_s,q,Q)$ and the coordinate periods,
Eq.~\eqref{eq:DeltaI} contains no explicit $\alpha$.  At a horizon or cap,
$f=0$, and differentiating the algebraic metric equation gives
Eq.~\eqref{eq:endpointderivatives}.  The endpoint constraint and its derivative
therefore contain no explicit $\alpha$.  After renormalization, the
state-dependent part of the gravitational contribution at the outer boundary
falls off as $O(\ln^2\rho/\rho^2)$.  The $\alpha$ dependence of the
phase diagram enters through $\Ba$ and $\ella$, which appear in the
boundary-period matching and in the relation between the reduced variables
and the physical boundary data.

The nonzero-charge coexistence curve in Fig.~\ref{fig:phaseboundary} is
obtained with the grand-canonical boundary condition $\cfin=1$, which fixes
the finite asymptotic coefficient $a_i$.  Within this static family, the
fixed-current condition in Eq.~\eqref{eq:logmodes} requires $q=Q=0$.  A global
phase diagram also requires all smooth branches and their fluctuation spectra.

 Black holes and solitons in 3DEGB  exhibit interesting phase behaviour
 that, while similar the situation  in (2+1)-dimensional Einstein Gravity,
offer greater scope for exploration of
a broader range of phenomena.  
 More complicated cases involving rotation would make an interesting future study.

\section*{Acknowledgements}
This work was supported in part by the Natural Sciences and Engineering Research Council of Canada.

\appendix
\section{Radial derivation of the Euclidean action}
\label{app:radial}

\renewcommand{\theequation}{A\arabic{equation}}
\setcounter{equation}{0}

We evaluate the on-shell action directly on the black-hole and soliton
solutions.

We obtain the boundary term required for a Dirichlet variational principle by
applying the general Horndeski result of Ref.~\cite{PadillaSivanesan2012} to
Eq.~\eqref{eq:Lphi}.  The relevant integration-by-parts identity is
\begin{equation}
-4\alpha\phi G^{ab}\nabla_a\nabla_b\phi
=4\alpha G^{ab}\nabla_a\phi\nabla_b\phi
-4\alpha\nabla_a\!\left(\phi G^{ab}\nabla_b\phi\right).
\end{equation}
When applying the Horndeski result to the original action, the boundary
contribution of the total derivative must be subtracted.
Since $\phi=\ln(r/\ell)$ is constant along each surface of constant $r$,
evaluating the resulting boundary term at $r=\rho$ and integrating over the
Euclidean time and angular coordinates gives
\begin{align}
I_{D,b}
&=-\frac{\beta_b\eta_b}{16\pi}
\left[
r f_b'+2f_b+\frac{4\alpha f_b^2}{3r^2}
\right]_{r=\rho},
\nonumber\\
I_{D,s}
&=-\frac{\beta_s\eta_s}{16\pi}
\left[
r f_s'+2f_s+\frac{4\alpha f_s^2}{3r^2}
\right]_{r=\rho}.
\end{align}
For the flat, homogeneous saddles considered here, the gravitational
counterterms that cancel the outer divergence are
\begin{align}
I_{g,\mathrm{ct},b}
&=\frac{\beta_b\eta_b}{8\pi}
\frac{\rho\sqrt{f_b(\rho)}}{\ella}
\left(1+\frac{2\alpha}{3\ella^2}\right),
\nonumber\\
I_{g,\mathrm{ct},s}
&=\frac{\beta_s\eta_s}{8\pi}
\frac{\rho\sqrt{f_s(\rho)}}{\ella}
\left(1+\frac{2\alpha}{3\ella^2}\right).
\end{align}

For either Euclidean metric, the radial reduction uses
\begin{equation}
R=-f''-\frac{2f'}{r},
\qquad
r\mathcal L_\phi
=-\left(\frac{f^2}{r^2}\right)'.
\end{equation}
Here $f=f_b$ for the black hole and $f=f_s$ for the soliton.  Integrating the
bulk action by parts and adding the boundary terms gives
\begin{align}
I_{g,b}
&=\frac{\beta_b\eta_b}{16\pi}\left\{
\left[
-f_b-\frac{r^2}{\ell^2}
-\frac{\alpha f_b^2}{3r^2}
+\frac{2r\sqrt{f_b}}{\ella}
\left(1+\frac{2\alpha}{3\ella^2}\right)
\right]_{r=\rho}
-\left[
r f_b'-\frac{r^2}{\ell^2}
\right]_{r=r_+}
\right\},
\nonumber\\
I_{g,s}
&=\frac{\beta_s\eta_s}{16\pi}\left\{
\left[
-f_s-\frac{r^2}{\ell^2}
-\frac{\alpha f_s^2}{3r^2}
+\frac{2r\sqrt{f_s}}{\ella}
\left(1+\frac{2\alpha}{3\ella^2}\right)
\right]_{r=\rho}
-\left[
r f_s'-\frac{r^2}{\ell^2}
\right]_{r=r_s}
\right\}.
\end{align}
The horizon and cap are smooth endpoints, so no separate inner-boundary term
is included.

Eq.~\eqref{eq:Balpha} implies
\begin{equation}
\frac{1}{\ell^2}
=\frac{1}{\ella^2}+\frac{\alpha}{\ella^4}.
\end{equation}
Using this relation and the asymptotic forms following
Eq.~\eqref{eq:Balpha}, the black-hole and soliton outer-boundary terms both
fall off as $O(\ln^2\rho/\rho^2)$.  At the smooth endpoints,
\begin{equation}
f_b'(r_+)=\frac{2r_+}{\ell^2}-\frac{2q^2}{r_+},
\qquad
f_s'(r_s)=\frac{2r_s}{\ell^2}+\frac{2Q^2}{r_s},
\label{eq:endpointderivatives}
\end{equation}
so the finite gravitational contributions are
\begin{align}
I_{g,b}
&=\frac{\beta_b\eta_b}{16\pi}
\left(-\frac{r_+^2}{\ell^2}+2q^2\right),
\nonumber\\
I_{g,s}
&=\frac{\beta_s\eta_s}{16\pi}
\left(-\frac{r_s^2}{\ell^2}-2Q^2\right).
\end{align}

The Maxwell action in Eq.~\eqref{eq:EuclideanMaxwellaction} gives
\begin{align}
I_{M,b}^{\mathrm{bulk}}
&=-\frac{\beta_b\eta_b}{8\pi}q^2\ln\frac{\rho}{r_+},
&
I_{M,s}^{\mathrm{bulk}}
&=\frac{\beta_s\eta_s}{8\pi}Q^2\ln\frac{\rho}{r_s}.
\end{align}
The opposite signs follow from the imaginary Euclidean electric potential.
Evaluating the counterterm in Eq.~\eqref{eq:Maxwellct} gives
\begin{align}
I_{M,\mathrm{ct},b}^{(\cfin)}
&=\frac{\beta_b\eta_b}{8\pi}q^2
\frac{\ella\sqrt{f_b(\rho)}}{\rho}
\left[\ln\frac{\rho}{r_0}+\cfin-1\right],
\nonumber\\
I_{M,\mathrm{ct},s}^{(\cfin)}
&=-\frac{\beta_s\eta_s}{8\pi}Q^2
\frac{\ella\sqrt{f_s(\rho)}}{\rho}
\left[\ln\frac{\rho}{r_0}+\cfin-1\right].
\end{align}
Taking $\rho\to\infty$, with $\ella\sqrt{f_b(\rho)}/\rho\to1$ and
$\ella\sqrt{f_s(\rho)}/\rho\to1$, gives
\begin{equation}
I_{M,b}^{(\cfin)}
=-\frac{\beta_b\eta_b}{8\pi}q^2\Lb,
\qquad
I_{M,s}^{(\cfin)}
=\frac{\beta_s\eta_s}{8\pi}Q^2\Ls.
\end{equation}

Adding the gravitational and Maxwell contributions gives
\begin{align}
I_b^{(\cfin)}
&=\frac{\beta_b\eta_b}{16\pi}
\left[-\frac{r_+^2}{\ell^2}+2q^2(1-\Lb)\right],
\nonumber\\
I_s^{(\cfin)}
&=\frac{\beta_s\eta_s}{16\pi}
\left[-\frac{r_s^2}{\ell^2}-2Q^2(1-\Ls)\right].
\end{align}
These expressions reproduce Eq.~\eqref{eq:individualactions}.  Using
$\beta_s\eta_s=\beta_b\eta_b$ from Eq.~\eqref{eq:periodmatch} then gives
Eq.~\eqref{eq:DeltaI}.

\bibliographystyle{unsrt}
\bibliography{ref}

\end{document}